\documentclass[twocolumn,showpacs,preprintnumbers,prb]{revtex4}
\usepackage{graphicx}
\usepackage{hyperref}

\begin{document}

\title{Thermodynamics of spin-$\frac{1}{2}$ tetrameric Heisenberg
antiferromagnetic chain}
\author{Shou-Shu Gong$^{1}$, Song Gao$^{2}$ and Gang Su$^{1,\ast}$}
\affiliation{$^{1}$College of Physical Sciences, Graduate University of Chinese Academy
of Sciences, P. O. Box 4588, Beijing 100049, People's Republic of China
\linebreak $^{2}$College of Chemistry and Molecular Engineering, State Key
Laboratory of Rare Earth Materials Chemistry and Applications, Peking
University, Beijing 100871, People's Republic of China}

\begin{abstract}
The thermodynamic properties of a spin $S$=$1/2$ tetrameric Heisenberg
antiferromagnetic chain with alternating interactions AF$_{1}$-AF$_{2}$-AF$%
_{1}$-F (AF and F denote the antiferromagnetic and ferromagnetic couplings,
respectively) are studied by means of the transfer-matrix renormalization
group method and Jordan-Wigner transformation. It is found that in the
absence of magnetic field, the thermodynamic behaviors are closely related
to the gapped low-lying excitations, and a novel structure with three peaks
in the temperature dependence of specific heat is unveiled. In a magnetic
field, a phase diagram in the temperature-field plane for the couplings
satisfying $J_{\mathrm{AF_{1}}}$=$J_{\mathrm{AF_{2}}}$=$J_{\mathrm{F}}$ is
obtained, in which various phases are identified. The temperature dependence
of thermodynamic quantities including the magnetization, susceptibility and
specific heat are studied to characterize the corresponding phases. It is
disclosed that the magnetization has a crossover behavior at low temperature
in the Luttinger liquid phase, which is shown falling into the same class as
that in the $S$=$1$ Haldane chain. In the plateau regime, the thermodynamic
behaviors alter at a certain field, which results from the crossing of two
excitation spectra. By means of the fermion mapping, it is uncovered that
the system has four spectra from fermion and hole excitations that are
responsible for the observed thermodynamic behaviors.
\end{abstract}

\maketitle

\section{Introduction}

In recent years, low-dimensional quantum magnets have received much
attention in condensed matter physics. In particular, one-dimensional
quantum spin chains with competing interactions that show exotic physical
properties have been extensively studied in the past decades. Among others,
the dimerized spin-$1/2$ antiferromagnetic (AF)-ferromagnetic (F) alternaing
chain has been widely studied both theoretically \cite{ABH,US,TB} and
experimentally,\cite{AF-FExp} where it was found that the system has a gap
from the singlet ground state to the triplet excited states, and can be
mapped onto the $S$=$1$ Haldane chain \cite{Hida,SWY,Zheng} if the F
couplings dominate. Moreover, a series of trimerized compounds, including
the $S$=$1/2$ antiferromagnetic F-F-AF chain of 3CuCl$_{2}$$\cdot 2$\textit{%
dx }(\textit{dx}=1,4-dioxane) \cite{F-F-AFExp} and the ferrimagnetic AF-AF-F
chains of [Mn(\textit{L}$_{2}$)(N3)$_{2}$]$_{n}$ (\textit{L}%
=3-methylpyridine) with $S$=$5/2$,\cite{MAM} [\textit{M}(4,4$^{\prime }$%
bipy)(N$_{3}$)$_{2}$]$_{n}$(bipy=bipyridine) with M=Co ($S$=$3/2$) and Ni ($S
$=$1$),\cite{SML} and [Mn(N$_{3}$)$_{2}$(bpee)]$_{n}$
(bpee=trans-1,2-bis(4-pyridyl)ethylene) with $S$=$5/2$,\cite{ADE} have been
synthesized in experiments. In such trimerized spin chains, the topological
quantization of magnetization, i.e., the magnetization plateau, has been
predicted theoretically.\cite{KO,AH,WC,Gu,Gong} The predictions were based
on the theorem proposed by Oshikawa, Yamanaka, and Affleck,\cite{OYA}
which extends the Lieb-Schultz-Mattis theorem to give a necessary condition
for the appearance of magnetization plateau in the spin chains with
translational symmetry. However, the plateau was not observed experimently
in the above-mentioned compounds owing to the weak AF couplings. A plateau
at $m$=$1/4$ ($m$ is the magnetization per site) was later observed in an $S$%
=$1/2$ tetrameric F-F-AF-AF ferrimagnet Cu(3-Clpy)$_{2}$(N$_{3}$)$_{2}$ \cite%
{ARM,MYK,SY,TS,HYL} due to the strong AF couplings.

Recently, a spin $S$=$1/2$ tetrameric Heisenberg antiferromagnetic chain
(HAFC) with AF$_{1}$-AF$_{2}$-AF$_{1}$-F interactions has been studied,
whose ground state was found to be in a gapped Haldane-like phase while,
importantly, it cannot be reduced to an integer spin chain.\cite{GSS} Twenty
six years ago, Haldane \cite{Haldane} conjectured that an isotropic HAFC
with an integer spin has a finite gap from the singlet ground state to the
triplet excited states, and the spin-spin correlation function decays
exponentially, while the HAFC with half-integer spin has a gapless spectrum
and a correlation function with a power-law decay. Although there is no
rigorous proof for a general case until now, Haldane's scenario has been
confirmed experimentally and numerically in many systems (e.g. Ref.%
\onlinecite{AKLT}). Besides the HAFCs with integer spin, Haldane gap has
also been found in $S$=$1/2$ spin ladders \cite{WNT,BR} and AF-F alternating
Heisenberg chains.\cite{Hida} This is because these spin-$1/2$ systems were
found to be reducible to an $S$=$1$ HAFC when the F couplings are dominantly
larger than the AF couplings. However, the present $S$=$1/2$ AF$_{1}$-AF$_{2}
$-AF$_{1}$-F tetrameric HAFC cannot be reduced to an integer spin chain even
if the F coupling dominates, and a gapped state with $m$=$0$ was observed.%
\cite{GSS} By using a dual transformation, the $Z_{2}$$\times $$Z_{2}$
hidden symmetry is disclosed to be fully broken, and the string order is
found non-vanishing in the ground state, further suggesting that this spin-$%
1/2$ tetrameric HAFC system belongs to the Haldane-like phase,\cite{NR,KT}
which extends the substance of Haldane's scenario, namely, the Haldane gap
can appear in certain spin half-integer chains. Apart from the gapped state
with $m$=$0$, a mangetization plateau at $m$=$1/4$ was also found in this
system.\cite{GSS} In the critical magnetic fields where the magnetization
curve is singular, quantum phase transitions (QPTs)\cite{QPT} may happen and
consequently, phase crossovers are expected at finite temperature.

As this spin-1/2 AF$_{1}$-AF$_{2}$-AF$_{1}$-F tetrameric HAFC exhibits many
interesting behaviors at zero temperature, a deeper investigation is still
quite necessary. In this paper, we shall go on elaborating the
thermodynamics of this system by means of the transfer-matrix
renormalization group (TMRG) as well as the Jordan-Wigner (JW)
transformation, with emphasis on the effects of the couplings and the
external magnetic field on thermodynamical properties of the system. The
possible magnetic phase diagram at finite temperature will be presented. The
low-lying excitations that are closely related to the observed thermodynamic
behaviors will also be discussed.

The Hamiltonian of the $S$=$1/2$ tetrameric HAFC with alternating couplings
AF$_{1}$-AF$_{2}$-AF$_{1}$-F in a longitudinal magnetic field is given by 
\begin{eqnarray}
H &=&\sum_{j=1}^{N}(J_{\mathrm{AF_{1}}}\mathbf{S}_{4j-3}\cdot \mathbf{S}%
_{4j-2}+J_{\mathrm{AF_{2}}}\mathbf{S}_{4j-2}\cdot \mathbf{S}_{4j-1} 
\nonumber \\
&+&J_{\mathrm{AF_{1}}}\mathbf{S}_{4j-1}\cdot \mathbf{S}_{4j}-J_{\mathrm{F}}%
\mathbf{S}_{4j}\cdot \mathbf{S}_{4j+1})-h\sum_{j=1}^{4N}S_{j}^{z},
\label{model
Hamiltonian}
\end{eqnarray}
where $J_{\mathrm{AF_{1,2}}}$ ($>$$0$) denote the AF couplings, $J_{\mathrm{F%
}}$ ($>$$0$) denotes the F coupling, and $h$ is the magnetic field. We take $%
J_{\mathrm{AF_{1}}}$ as the energy scale and $g\mu_{B}$=$1$ for convenience. 
$N$ is the total number of the unit cells. The Boltzmann constant is taken
as $k_{B}$=$1$.

The numerical algorithm TMRG method,\cite{TMRG1,TMRG2,TMRG3} which is a
powerful tool for studying the thermodynamics of one-dimensional quantum
systems, will be primarily employed in the following investigations. As the
TMRG technique has been discussed in many reviews, we shall not repeat the
details here. In the following calculations, the width of the imaginary time
slice is taken as $\varepsilon $=$0.1$, and the error caused by the
Trotter-Suzuki decomposition is less than $10^{-3}$. During the TMRG
iterations, $80$ states are retained and the temperature is down to $T$=$%
0.02J_{\mathrm{AF_{1}}}$ in general. In the Luttinger liquid and gapless
phases, the temperature is down lower than $0.01J_{\mathrm{AF_{1}}}$ when
calculating the magnetization and susceptibility. The truncation error is
less than $10^{-4}$ in all calculations.

The other parts of the paper are organized as follows. In Sec. \uppercase%
\expandafter{\romannumeral2}, we shall present the TMRG results of the
thermodynamic quantities in the absence of magnetic field. In Sec. \uppercase%
\expandafter{\romannumeral3}, a magnetic phase diagram at finite temperature
will be proposed for the case with $J_{\mathrm{AF_{1}}}$=$J_{\mathrm{AF_{2}}}
$=$J_{\mathrm{F}}$, and the thermodynamic properties will be discussed in
the various phases. In Sec. \uppercase\expandafter{\romannumeral4}, we shall
invoke the mean-field results from the JW transformation to explain the
behaviors observed in Secs. \uppercase\expandafter{\romannumeral2} and %
\uppercase\expandafter{\romannumeral3}. Finally, a summary and discussion
will be given.

\section{Zero-field thermodynamic properties}

\subsection{Specific heat}

\begin{figure}[tbp]
\includegraphics[width=1.0\linewidth,clip]{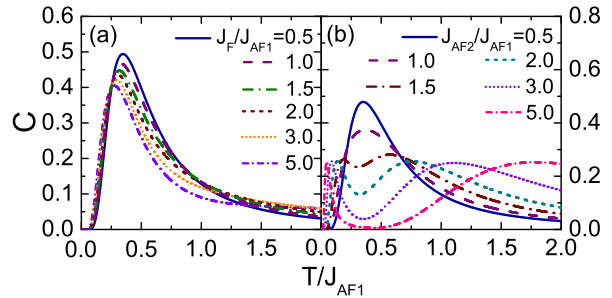}
\caption{(Color online) Temperature dependence of the specific heat of the
isolated tetramer systems with (a) $J_{\mathrm{AF_{2}}}$=$0$; and (b) $J_{%
\mathrm{F}}$=$0$.}
\label{Cv1}
\end{figure}

Let us first look at the temperature dependence of the specific heat for the
isolated tetramer systems with $J_{\mathrm{AF_{2}}}$=$0$ or $J_{\mathrm{F}}$=%
$0$. For $J_{\mathrm{AF_{2}}}$=$0$, the specific heat decays exponentially
as $T$$\rightarrow $$0$ and has a sharp peak at low temperature that shifts
slightly to lower temperatures with increasing $J_{\mathrm{F}}$, as shown in
Fig. \ref{Cv1}(a). For $J_{\mathrm{F}}$=$0$, the specific heat has a single
peak when $J_{\mathrm{AF_{2}}}$/$J_{\mathrm{AF_{1}}}$$<$$1$. If $J_{\mathrm{%
AF_{2}}}$ exceeds $J_{\mathrm{AF_{1}}}$, with increasing $J_{\mathrm{AF_{2}}%
} $ the single peak splits into double peaks, one of which moves to lower
temperatures while the other moves to higher temperatures, as shown in Fig. %
\ref{Cv1}(b). The distinct behaviors of the specific heat for the two
tetramer systems are owing to their different energy spectra. When $J_{%
\mathrm{AF_{2}}}$=$0$, for each tetramer there are four energy levels that
decrease with increasing $J_{\mathrm{F}}$, while the gaps among them vary
slightly, yielding a small shift of the peak. When $J_{\mathrm{F}}$=$0$,
with increasing $J_{\mathrm{AF_{2}}}$ the gap between the ground state and
the first excited states diminishes rapidly, accounting for the shift of the
low-temperature peak, while the shift of the high-temperature peak is owing
to the enlarged gaps between the energy levels. These limiting cases offer
useful information for better understanding the features of the specific
heat of the systems with arbitrary couplings, which will be discussed below. 

\begin{figure}[tbp]
\includegraphics[width=1.0\linewidth,clip]{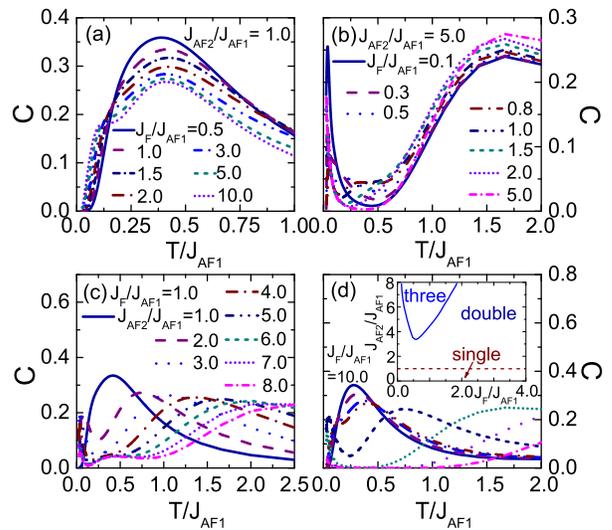}
\caption{(Color online) Temperature dependence of the specific heat of the
tetrameric chain for (a) $J_{\mathrm{AF_{2}}}/J_{\mathrm{AF_{1}}}$=$1.0$;
(b) $J_{\mathrm{AF_{2}} }/J_{\mathrm{AF_{1}}}$=$5.0$; (c) $J_{\mathrm{F}}/J_{%
\mathrm{AF_{1}}}$=$1.0$; and (d) $J_{\mathrm{F}}/J_{\mathrm{AF_{1}}}$=$10.0$
with $J_{\mathrm{AF_{2}}}/J_{\mathrm{AF_{1}}}$=$0.5,0.7,0.9,1.0, 2.0,5.0,8.0$
and $12.0$ from top to bottom. The inset of (d) shows the parameter regions
where the specific heat has different peak structures.}
\label{Cv2}
\end{figure}

The effects of $J_{\mathrm{F}}$ on the specific heat are firstly discussed.
In Fig. \ref{Cv2}(a), the specific heat of the system with $J_{\mathrm{AF_{2}%
}}$/$J_{\mathrm{AF_{1}}}$=$1.0$ are presented for different $J_{\mathrm{F}}$%
. The specific heat has a single peak and decays exponentially as $T$$%
\rightarrow $$0$, indicating a gapped excitation. With increasing $J_{%
\mathrm{F}}$, the single peak declines and a shoulder appears to be
gradually prominent at low temperatures. Thus, it is expected that the
system has at least two gapped excitations. With increasing $J_{\mathrm{F}}$%
, the gap that is related to the lower excitation decreases slowly,\cite{GSS}
yielding the emergence of the shoulder. Figure \ref{Cv2}(b) shows the
specific heat of the system with $J_{\mathrm{AF_{2}}}$/$J_{\mathrm{AF_{1}}}$=%
$5.0$ for different $J_{\mathrm{F}}$. As $J_{\mathrm{AF_{2}}}$ is large, the
specific heat has two peaks for $J_{\mathrm{F}}$/$J_{\mathrm{AF_{1}}}$=$0.1 $%
, like the tetramer system with $J_{\mathrm{F}}$=$0$. With increasing $J_{%
\mathrm{F}}$, a novel peak emerges between the low and high temperature
peaks, which shifts to higher temperatures with further increase of the F
coupling until it is merged into the high temperature peak. It can be seen
that the specific heat behaves rather differently with $J_{\mathrm{F}}$ for
different $J_{\mathrm{AF_{2}}}$, and in certain couplings region, the
specific heat can exhibit a novel three-peak structure.

Next, we study the effects of $J_{\mathrm{AF_{2}}}$ on the specific heat. In
Fig. \ref{Cv2}(c), the specific heat of the system with $J_{\mathrm{F}}$/$J_{%
\mathrm{AF_{1}}}$=$1.0$ are plotted for various $J_{\mathrm{AF_{2}}}$. When $%
J_{\mathrm{AF_{2}}}$=$J_{\mathrm{AF_{1}}}$, the specific heat has a single
peak and decays exponentially as $T$$\rightarrow $$0$. When $J_{\mathrm{%
AF_{2}}}$ exceeds $J_{\mathrm{AF_{1}}}$, the single peak splits into double
peaks, one of which shifts to lower temperature side while another moves to
the higher temperature side with increasing $J_{\mathrm{AF_{2}}}$. These
behaviors could be understood by means of the corresponding tetramer system
where the gaps between the energy levels are decreasing rapidly with
increasing $J_{\mathrm{AF_{2}}}$.\cite{GSS} It is noticed that the novel
peak emerges when $J_{\mathrm{AF_{2}}}$ is large enough, e.g. $J_{\mathrm{%
AF_{2}}}$/$J_{\mathrm{AF_{1}}}$=$4.0$ for $J_{\mathrm{F}}$/$J_{\mathrm{AF_{1}%
}}$=$1.0$. When $J_{\mathrm{AF_{2}}}$ continues to increase, the novel peak
is nearly invariant. When $J_{\mathrm{F}}$ is too large, as shown in Fig. %
\ref{Cv2}(d) for $J_{\mathrm{F}}$/$J_{\mathrm{AF_{1}}}$=$10.0$, the novel
peak is absent. These results indicate that the novel peak is yielded by
increasing $J_{\mathrm{F}} $ after the double peaks have been induced by large $J_{%
\mathrm{AF_{2}}}$.

The inset of Fig. \ref{Cv2}(d) shows the parameter regions where the
specific heat has different peak structures. The dashed line separates the
regions of the single peak and double peaks, while the solid line separates
the regions of the double and three peaks. The observations in Figs. \ref%
{Cv2}(a)-(c) are manifested clearly in this depiction. It is shown that when 
$J_{\mathrm{AF_{2}}}$$>$$J_{\mathrm{AF_{1}}}$, the single peak starts to
split into double peaks. When $J_{\mathrm{AF_{2}}}/J_{\mathrm{AF_{1}}}$
exceeds about $3.5$, the novel peak can appear with increasing $J_{\mathrm{F}%
}$, and it would merge into the high temperature peak with further increase
of the coupling. As the intermediate region is enlarged by increasing $J_{%
\mathrm{AF_{2}}}$, the parameter region of $J_{\mathrm{F}}$ for the
emergence of the novel peak is wider for larger $J_{\mathrm{AF_{2}}}$. It is
interesting to point out that the emergence of three peaks of the specific
heat in the absence of magnetic field is nontrivial, which is not the usual
feature in low-dimensional quantum magnets.

\subsection{Susceptibility}

\begin{figure}[tbp]
\includegraphics[width=1.0\linewidth,clip]{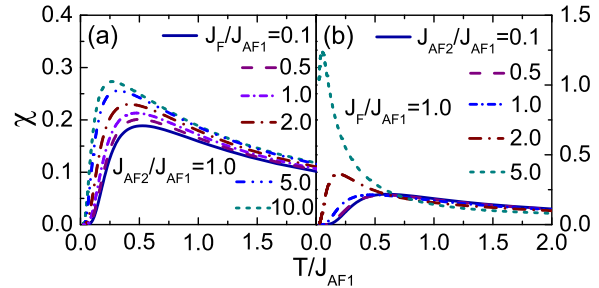}
\caption{(Color online) Temperature dependence of the spin susceptibility of
the tetrameric chain for (a) $J_{\mathrm{AF_{2}}}$/$J_{\mathrm{AF_{1}}}$=$%
1.0 $ with different $J_{\mathrm{F}}$; and (b) $J_{\mathrm{F}}$/$J_{\mathrm{%
AF_{1}}}$=$1.0$ with different $J_{\mathrm{AF_{2}}}$.}
\label{Susceptibility}
\end{figure}

The temperature dependence of the susceptibility is presented in Fig. \ref%
{Susceptibility} for different cases. It is shown that the susceptibility
has a peak and decreases exponentially as $T$$\rightarrow $$0$. With
increasing $J_{\mathrm{F}}$ or $J_{\mathrm{AF_{2}}}$, the peak 
shifts to lower temperatures with the height enhanced, which
are consistent with the diminution of the gap and the behaviors of the
specific heat. As the gap decreases slowly with $J_{\mathrm{F}}$ but rapidly
with $J_{\mathrm{AF_{2}}}$,\cite{GSS} the susceptibility changes slightly
with $J_{\mathrm{F}}$ but dramatically with $J_{\mathrm{AF_{2}}}$, as
demonstrated in Figs. \ref{Susceptibility}(a) and (b), respectively. At high
temperatures, the gap is suppressed by thermal fluctuations, and the
susceptibility goes to coincidence for different couplings. These behaviors
imply the distinctions of the low-lying excitations, which will be
reexamined in terms of the spinless fermions in Sec. \uppercase%
\expandafter{\romannumeral4}.

\section{Thermodynamics in magnetic fields}

In this section, the thermodynamic properties of the system with $J_{\mathrm{%
AF_{1}}}$=$J_{\mathrm{AF_{2}}}$=$J_{\mathrm{F}}$ in the presence of a
magnetic field are studied by means of the TMRG method. A phase diagram in the
temperature-field plane is proposed, and the magnetization, susceptibility,
and specific heat are investigated accordingly in the various phases.

\subsection{Phase diagram}

\begin{figure}[tbp]
\includegraphics[width=0.7\linewidth,clip]{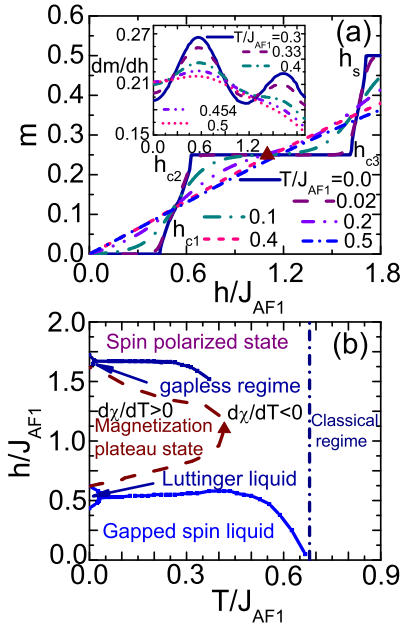}
\caption{(Color online) (a) Magnetization curves at different temperatures.
The inset shows $\partial{m}/\partial{h}$ as a function of $h$ at different
temperatures. (b) Phase diagram in the temperature-field plane. The solid
lines and the dashed line are obtained by observing the peaks of $\partial
m/\partial h$ and the points with $\partial \protect\chi/\partial T$=$0$,
respectively.}
\label{phase}
\end{figure}

As shown in Fig. \ref{phase}(a), the zero-temperature magnetization curve is
singular at the critical fields $h_{c_{1}},h_{c_{2}},h_{c_{3}}$ and $h_{s}$,%
\cite{GSS} suggesting that QPTs\cite{QPT} may happen. These transitions are
measured by the divergent peaks of $\partial {m}/\partial {h}$, which
separate the Haldane-like phase, Luttinger liquid (LL) phase, plateau phase,
gapless phase, and polarized state of the system in a magnetic field. At
finite temperature, the magnetization plateaus are smeared out, and the
peaks of $\partial {m}/\partial {h}$ become analytic, which, however, can
still describe the crossover behaviors of the various phases. Therefore, the
magnetization process at different temperatures will be studied to obtain
the phase diagram in the temperature-field plane.

Figure \ref{phase}(a) shows that with increasing temperature, the two peaks
of $\partial {m}/\partial {h}$ at $h_{c_{1}}$ ($h_{c_{3}}$) and $h_{c_{2}}$ (%
$h_{s}$) gradually merge into a single peak at rather low temperature,
indicating the crossovers from the LL (gapless) regime to other regimes.
With further increasing temperature, the high-field peak that separates the
plateau and spin polarized regimes disappears at $T$$\approx$$0.44J_{\mathrm{%
AF_{1}}}$, and the low-field peak that separates the gapped spin liquid and
plateau regimes disappears at $T$$\approx$$0.68J_{\mathrm{AF_{1}}}$ [the
inset of Fig. \ref{phase}(a)]. The shifts of the peaks in the
temperature-field plane compose of the crossover lines. Based on the
observations, we propose a phase diagram in the temperature-field plane as
shown in Fig. \ref{phase}(b), from which one may observe that the system has
the gapped spin liquid, LL, magnetization plateau, gapless, spin polarized,
and classical phases.

\subsection{Magnetization}

\begin{figure}[tbp]
\includegraphics[width=1.0\linewidth,clip]{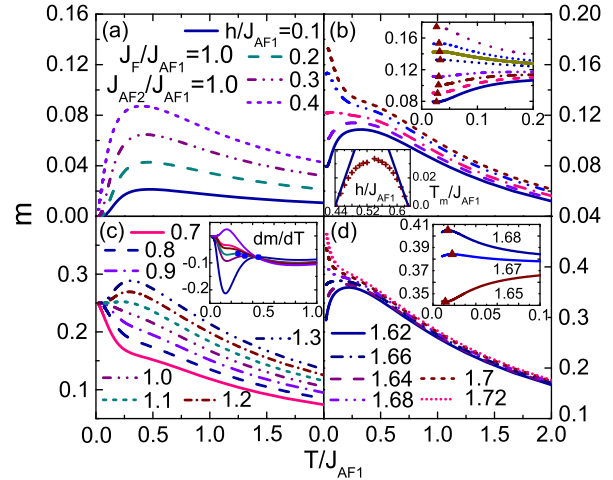}
\caption{(Color online) Magnetization as a function of temperature for (a) $%
h $$<$$h_{c_{1}}$; (b) $h_{c_{1}}$$<$$h$$<$$h_{c_{2}}$; (c) $h_{c_{2}}$$<$$h$%
$<$$h_{c_{3}}$; and (d) $h_{c_{3}}$$<$$h$$<$$h_{s}$. The insets show (b) the
minima (maxima) and the field dependence of the crossover temperature; (c)
the behavior of $\partial{m}/\partial{T}$ for $h$/$J_{\mathrm{AF_{1}}}$=$%
0.9,1.0,1.02,1.04,1.05$ and $1.1$ from bottom to top; and (d) the minima and
maxima in the gapless regime.}
\label{Mz}
\end{figure}

The temperature dependence of the magnetization $m(T)$ in the various
regimes are investigated. When $h$$<$$h_{c_{1}}$, the low-lying excitation
is gapped. With decreasing temperature, the magnetization $m$ first
increases with a power law, then goes down, and finally decays exponentially
to zero with the effective gap $\Delta _{\mathrm{eff}}$=$h_{c_{1}}-h$ as $T$$%
\rightarrow $$0$. As expected, with decreasing the effective gap, the peak
of $m(T)$ moves to lower temperature side with the amplitude enhanced, as
shown in Fig. \ref{Mz}(a).

In the LL regime when $h_{c_{1}}$$<$$h$$<$$h_{c_{2}}$, the gap is closed by
the field, and the system undergoes a commensurate-incommensurate transition
at $h$=$h_{c_{1}}$ in the ground state.\cite{CG} At finite temperatures, $%
m(T)$ shows a minimum or maximum at low temperatures, as shown in Fig. \ref%
{Mz}(b). The minima are close to the crossover boundary between the gapped
spin liquid and LL regimes, while the maxima are close to that between the
magnetization plateau and LL regimes, indicating the nonsingular crossovers
from the LL to the high-temperature regimes. Maeda, Hotta, and Oshikawa \cite%
{MHO} pointed out that this crossover is universal in general gapped
one-dimensional (1D) spin systems with axial symmetry, resulting from the
minus derivative of the density of states near the critical field and the
variation of the Fermi velocity $v_{F}$. In Sec. \uppercase%
\expandafter{\romannumeral4}, this observation would be reproduced by the JW
transformation and interpreted more physically.

In Ref. \onlinecite{MHO}, a linear dependence of the transition temperature (%
$T_{m}$) on the field $T_{m}$=$x_{0}(h-h_{c})$ ($h_{c}$ is the critical
field where the gap is closed) near the critical field $h_{c}$ was proposed
for the $S$=$1$ Haldane chain. The field dependence of the crossover
temperature for the present tetrameric HAFC is shown in the lower inset of
Fig. \ref{Mz}(b). It can be seen that the crossover temperature $T_{m}$
varies like a sine function of the field and behaves linearly near the
critical fields $h_{c_{1}}$ and $h_{c_{2}}$. The coefficient $x_{0}$$\simeq $%
$0.76238$ proposed in Ref. \onlinecite{MHO} for the $S$=$1$ Haldane chain
near $h_{c}$ fits well to our data near both $h_{c_{1}}$ and $h_{c_{2}}$
(see the solid lines in the inset), indicating that this tetrameric chain
slightly above $h_{c_{1}}$ and below $h_{c_{2}}$ could also be well
described by the free fermion theory,\cite{IA2,HJS,LL} and the crossover in
this LL phase is of the same class as that in the Haldane chain. Although
this tetrameric chain cannot be reduced to a typical Haldane chain with an
integer spin, their analogous dispersion relation of the low-lying
excitations are revealed by the same class of this crossover behavior of magnetization.

In the magnetization plateau, $m$ approaches $0.25$ as $T$$\rightarrow $$0$.
As shown in Fig. \ref{Mz}(c), when $h_{c_{2}}$$<$$h$$<$$h_{m}$$\simeq 1.1J_{%
\mathrm{AF_{1}}}$, $m$ increases slowly with cooling temperature to a
certain value and, then, rises rapidly to $0.25$, where $h_{m}$ is the
crossing field of the magnetization curves at low temperatures in the
plateau states, as marked by the triangle in Fig. \ref{phase}(a). When $h_{m}
$$<$$h$$<$$h_{c_{3}}$, $m$ increases to a maximum and then declines to $0.25$.
The different behaviors are also observed in the magnetization curves at
various temperatures [Fig. \ref{phase}(a)]. In the field $h_{m}$, $m$ changes slowly at low
temperature, like the $m(T)$ curve with $h$/$J_{\mathrm{AF_{1}}}$=$1.1$ in
Fig. \ref{Mz}(c). These magnetic behaviors in the plateau state have also
been noted in the spin-$1/2 $ trimerized\cite{Gu} and F-F-AF-AF tetrameric
chains.\cite{MYK} In Sec. \uppercase\expandafter{\romannumeral4}, it would
be found that these common features in the plateau states may result from
the crossing of fermion and hole excitations. The field $h_{m}$ is in the
middle of the plateau and corresponds to the midpoint of the gap between the
crossing fermion and hole spectra. When $h$$>$$h_{m}$, the maximum of $m(T)$
with $\partial m/\partial T$=$0$ measures the change of the magnetic
properties. For $h$$<$$h_{m}$, although $m$ keeps declining, such a change
also exists, which can be visible in the temperature dependence of $\partial
m/\partial T$. Corresponding to $\partial m/\partial T$=$0$ for $h$$>$$h_{m}$%
, $\partial m/\partial T$ has maxima that are marked by solid squares to
separate the high temperature regime for $h$$<$$h_{m}$, as shown in the
inset of Fig. \ref{Mz}(c). As the field approaches $h_{m}$, the maximum
tends to disappear.

In the gapless regime, the gap in the plateau state can be closed by
increasing the field. Thus, the magnetic behavior $m(T)$ is analogous to
that in the LL phase, where there are also minimum and maximum observed at
rather low temperatures [Fig. \ref{Mz}(d)]. Such a crossover in the gapless
phase between a $m$$\neq$$0$ plateau and the saturated state has not been
reported. The magnetic behaviors of the spin-$1/2$ F-F-AF antiferromagnetic
chain \cite{Gu} have been studied numerically, but no such a crossover is
observed between the $m$=$1/6$ plateau and the saturated state when the
temperature is down to $0.025J_{\mathrm{F}}$. It is also noticed that the
crossover temperatures in the gapless regime are lower than those in the LL
phase, which would be interpreted in terms of the spinless fermion in Sec. %
\uppercase\expandafter{\romannumeral4}.

\subsection{Susceptibility}

\begin{figure}[tbp]
\includegraphics[width=1.0\linewidth,clip]{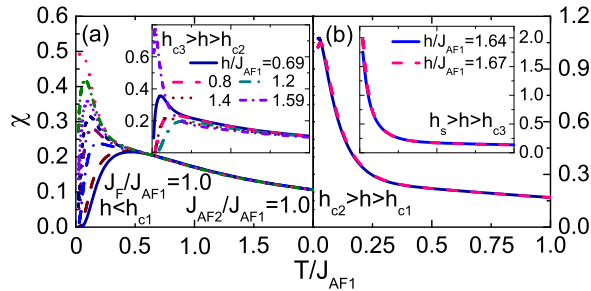}
\caption{(Color online) Temperature dependence of the susceptibility for (a) 
$h$/$J_{\mathrm{AF_{1}}}$=$0.1,0.2,0.3,0.34,0.36,0.38,0.4$ and $0.42$ from
bottom to top; (b) $h$/$J_{\mathrm{AF_{1}}}$=$0.5$ and $0.55$; the inset of
(a) $h_{c_{3}}$$>$$h$$>$$h_{c_{2}}$; and (b) $h_{s}$$>$$h$$>$$h_{c_{3}}$.}
\label{susceptibility2}
\end{figure}

The behaviors of the susceptibility $\chi $ in the various regimes will be
discussed in this subsection. When $h/J_{\mathrm{AF_{1}}}$ is less than
about $0.3$ in the gapped spin liquid, $\chi $ has a single peak at $T$$%
\simeq $$\Delta $ ($\Delta $ is the gap in the absence of magnetic field),
and approaches zero exponentially as $T$$\rightarrow $$0$. With further
increasing $h$, a new peak emerges and moves to lower temperatures with the
height enhanced, while the high temperature peak becomes smoother, as shown
in Fig. \ref{susceptibility2}(a). In the high temperature region $T$$>$$%
\Delta $, the susceptibility under different fields coincide because the gap
is suppressed by thermal fluctuations. In the fermion mapping in Sec. %
\uppercase\expandafter{\romannumeral4}, the system has two positive energy
excitations (fermion excitations) and two minus energy excitations (hole
excitations) when $h$=$0$. With the decrease of the excitation energies
induced by increasing the field, the susceptibility contributed from the
fermion excitations moves to lower temperatures with the amplitude enhanced,
while that from the hole excitations shifts to higher temperatures with the
height decreased, both of which are responsible for the behaviors observed
in Fig. \ref{susceptibility2}(a). In the LL, $\chi $ is finite as $T$$%
\rightarrow $$0 $, as shown in Fig. \ref{susceptibility2}(b). With cooling
temperature, $\chi $ increases slowly to the temperature $T$$\simeq $$\Delta 
$ and then has a sharp rise until to a peak at rather low temperatures,
which results from the closure of the gap.

In the plateau state, due to the open of a gap, the susceptibility has a
single peak, and approaches zero exponentially as $T$$\rightarrow $$0$. With
increasing the field, the peak moves to higher temperatures with the height
declined when $h$$<$$h_{m}$, while it shifts to lower temperatures with the
height enhanced after $h$ exceeds $h_{m}$, as shown in the inset of Fig. \ref%
{susceptibility2}(a). The field dependence of the peak temperature is
displayed by the dashed line in Fig. \ref{phase}(b), where the field with
the highest peak temperature is $h_{s}$, as marked by a triangle. These
distinct behaviors of the susceptibility also result from the crossing of
the fermion and hole spectra. When $h$$<$$h_{m}$, $\chi $ is dominated by a
hole branch whose gap is enhanced with increasing the field, yielding the
peak to move to higher temperatures with the height decreased. After $h$
exceeds $h_{m}$, the two spectra cross and the susceptibility is dominated
by the fermion branch whose gap declines with the increasing field, yielding
the peak to move to lower temperatures with the height enhanced. Different
from the gapped spin liquid, the susceptibility does not show double peaks
in the plateau phase. In the gapless regime, owing to the closure of the
gap, the susceptibility exhibits the same features as those in the LL, as
shown in the inset of Fig. \ref{susceptibility2}(b).

\subsection{Specific heat}

\begin{figure}[tbp]
\includegraphics[width=1.0\linewidth,clip]{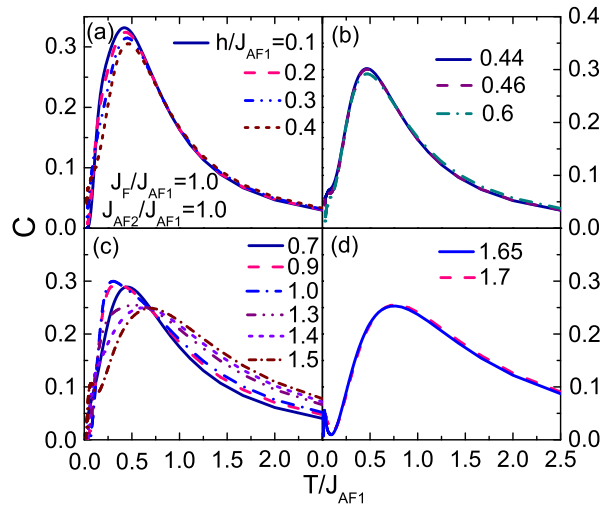}
\caption{(Color online) Specific heat of the system with (a) $h_{c_{1}}$$>$$%
h $; (b) $h_{c2}$$>$$h$$>$$h_{c1}$; (c) $h_{c3}$$>$$h$$>$$h_{c2}$; and (d) $%
h_{s}$$>$$h$$>$$h_{c3}$.}
\label{Cv3}
\end{figure}

In this subsection, the specific heat is explored in detail. When $h$$<$$%
h_{c_{1}}$, the specific heat $C$ has a single peak and approaches zero
exponentially as $T$$\rightarrow $$0$. The peak shifts to higher
temperatures with the height decreased when $\Delta _{\mathrm{eff}}$
diminishes [Fig. \ref{Cv3}(a)]. This shift is attributed to the hole
excitations whose gaps increase with increasing the field. It is noticed
that the peak temperature, $T_{peak}$/$J_{\mathrm{AF_{1}}}$$\simeq $$\Delta $%
, in this system is distinct from the result $T_{peak}/J$$\simeq$$2\Delta
^{\prime }$ of the $S$=$1$ Haldane chain ($J$ and $\Delta ^{\prime }$ are
the coupling and gap of the $S$=$1$ Haldane chain, respectively).\cite{YM}
As the field $h$ approaches $h_{c_{1}}$, a shoulder gradually emerges at low
temperature, which is a signature of approaching the quantum critical point 
\cite{SS,QPT} and is from the hole excitations. 

In the LL, the linear temperature dependence of the specific heat at low
temperature is observed. With a further increase of the field, the system
shifts away from the quantum critical point $h_{c_{1}}$. Thus, the shoulder
at low temperature is smoothed down gradually, as shown in Fig. \ref{Cv3}%
(b). The disappearance of the shoulder is attributed to the hole excitations.

In the plateau state, the specific heat decays exponentially when $T$$%
\rightarrow $$0$ because of the open of a gap. Near the lower critical field 
$h_{c_{2}}$, the shoulder that emerges in the LL vanishes gradually, while
near the upper critical field $h_{c_{3}}$, a double-peak structure emerges.
As expected, the crossing of the fermion and hole spectra affects the
behavior of the specific heat. With increasing the field, the peak of the
specific heat moves to lower temperatures when $h$$<$$h_{m}$, and when $h$
exceeds $h_{m}$, the peak starts to move to higher temperatures, as shown in
Fig. \ref{Cv3}(c). However, different from the magnetization and
susceptibility, the behavior of the specific heat cannot be characterized
simply only by the hole ($h$$<$$h_{m}$) or fermion ($h$$>$$h_{m}$)
excitations, although the crossing indeed changes the features of the
behavior. This is because the magnetization and susceptibility are
determined only by the occupied number of the excitations but the specific
heat is affected by both the numbers and energies of the quasiparticles.

In the gapless regime, with increasing the field, the high temperature peak
of the specific heat keeps nearly intact, while the low-temperature peak
that occurs near the critical field $h_{c_{3}}$ in the plateau state moves
to lower temperatures with the height declined [Fig. \ref{Cv3}(d)]. These
behaviors will be analyzed in the next section.

\section{Jordan-Wigner transformation and spinless fermion mapping}

In order to explain the thermodynamic behaviors observed in the above
sections, the elementary excitations of the system are studied using the JW
transformation.\cite{JW} The $S$=$1/2$ spin operators can be transformed
into the spinless fermion operators through JW transformation 
\begin{equation}
S_{i}^{+}=c_{i}^{\dagger }e^{i\pi \sum_{j<i}c_{j}^{\dagger }c_{j}},\quad
S_{i}^{z}=(c_{i}^{\dagger }c_{i}-\frac{1}{2}),
\end{equation}%
where $c_{i}^{\dagger }$ and $c_{i}$ are the creation and annihilation
operators of the spinless fermion, respectively. For this $S$=$1/2$
tetrameric HAFC, four kinds of spinless fermions should be introduced: 
\begin{eqnarray}
S_{4j-3}^{+} &=&a_{j}^{\dagger }exp[i\pi \sum_{m<j}(a_{m}^{\dagger
}a_{m}+b_{m}^{\dagger }b_{m}+c_{m}^{\dagger }c_{m}+d_{m}^{\dagger }d_{m})], 
\nonumber \\
S_{4j-2}^{+} &=&b_{j}^{\dagger }exp\{i\pi \lbrack \sum_{m<j}(a_{m}^{\dagger
}a_{m}+b_{m}^{\dagger }b_{m}+c_{m}^{\dagger }c_{m}+d_{m}^{\dagger }d_{m}) 
\nonumber \\
&+&a_{j}^{\dagger }a_{j}]\},  \nonumber \\
S_{4j-1}^{+} &=&c_{j}^{\dagger }exp\{i\pi \lbrack \sum_{m<j}(a_{m}^{\dagger
}a_{m}+b_{m}^{\dagger }b_{m}+c_{m}^{\dagger }c_{m}+d_{m}^{\dagger }d_{m}) 
\nonumber \\
&+&a_{j}^{\dagger }a_{j}+b_{j}^{\dagger }b_{j}]\},  \nonumber \\
S_{4j}^{+} &=&d_{j}^{\dagger }exp\{i\pi \lbrack \sum_{m<j}(a_{m}^{\dagger
}a_{m}+b_{m}^{\dagger }b_{m}+c_{m}^{\dagger }c_{m}+d_{m}^{\dagger }d_{m}) 
\nonumber \\
&+&a_{j}^{\dagger }a_{j}+b_{j}^{\dagger }b_{j}+c_{j}^{\dagger }c_{j}]\}, 
\nonumber
\end{eqnarray}%
\begin{eqnarray}
S_{4j-3}^{z} &=&a_{j}^{\dagger }a_{j}-\frac{1}{2},\quad
S_{4j-2}^{z}=b_{j}^{\dagger }b_{j}-\frac{1}{2},  \nonumber \\
S_{4j-1}^{z} &=&c_{j}^{\dagger }c_{j}-\frac{1}{2},\quad
S_{4j}^{z}=d_{j}^{\dagger }d_{j}-\frac{1}{2}.
\end{eqnarray}%
After the JW transformation, the $XY$ interactions of the original
Hamiltonian are transformed to the nearest-neighbor hoppings of the
fermions, and the Ising terms become the nearest-neighbor density-density
interactions that will be treated by the Hartree-Fock (HF) approximation. By
performing a cumbersome derivation, we obtain a mean-field Hamiltonian after
omitting the constant 
\begin{eqnarray}
H_{HF} &=&\sum_{j=1}^{N}\{\frac{1}{2}[J_{\mathrm{AF_{1}}}(d_{b}-\frac{1}{2}%
)-J_{\mathrm{F}}(d_{d}-\frac{1}{2})]a_{j}^{\dagger }a_{j}  \nonumber \\
&+&\frac{1}{2}[J_{\mathrm{AF_{1}}}(d_{a}-\frac{1}{2})+J_{\mathrm{AF_{2}}%
}(d_{c}-\frac{1}{2})]b_{j}^{\dagger }b_{j}  \nonumber \\
&+&\frac{1}{2}[J_{\mathrm{AF_{2}}}(d_{b}-\frac{1}{2})+J_{\mathrm{AF_{1}}%
}(d_{d}-\frac{1}{2})]c_{j}^{\dagger }c_{j}  \nonumber \\
&+&\frac{1}{2}[J_{\mathrm{AF_{1}}}(d_{c}-\frac{1}{2})-J_{\mathrm{F}}(d_{a}-%
\frac{1}{2})]d_{j}^{\dagger }d_{j}  \nonumber \\
&+&J_{\mathrm{AF_{1}}}(\frac{1}{2}-p_{AB})a_{j}^{\dagger }b_{j}+J_{\mathrm{%
AF_{2}}}(\frac{1}{2}-p_{BC})b_{j}^{\dagger }c_{j}  \nonumber \\
&+&J_{\mathrm{AF_{1}}}(\frac{1}{2}-p_{CD})c_{j}^{\dagger }d_{j}-J_{\mathrm{F}%
}(\frac{1}{2}-p_{DA})d_{j}^{\dagger }a_{j+1}  \nonumber \\
&+&h.c.\}-h\sum_{j=1}^{N}(a_{j}^{\dagger }a_{j}+b_{j}^{\dagger
}b_{j}+c_{j}^{\dagger }c_{j}+d_{j}^{\dagger }d_{j}),  \label{HFH}
\end{eqnarray}%
where the occupied fermion numbers are $d_{a}$=$\langle a_{j}^{\dagger
}a_{j}\rangle $, $d_{b}$=$\langle b_{j}^{\dagger }b_{j}\rangle $, $d_{c}$=$%
\langle c_{j}^{\dagger }c_{j}\rangle $, $d_{d}$=$\langle d_{j}^{\dagger
}d_{j}\rangle $, and the covalent bondings are $p_{AB}$=$\langle
b_{j}^{\dagger }a_{j}\rangle $, $p_{BC}$=$\langle c_{j}^{\dagger
}b_{j}\rangle $, $p_{CD}$=$\langle d_{j}^{\dagger }c_{j}\rangle $, and $%
p_{DA}$=$\langle a_{j+1}^{\dagger }d_{j}\rangle $. The brackets $\langle
\cdots \rangle $ denote either the HF ground state average ($T$=$0$) or the
thermal average ($T$$\neq $$0$). By making the Fourier transform and then
Bogoliubov transformations, a quadratic Hamiltonian can be obtained 
\begin{equation}
H_{HF}=\sum_{k}(\omega _{k}^{\alpha }\alpha _{k}^{\dagger }\alpha
_{k}+\omega _{k}^{\beta }\beta _{k}^{\dagger }\beta _{k}+\omega _{k}^{\gamma
}\gamma _{k}^{\dagger }\gamma _{k}+\omega _{k}^{\lambda }\lambda
_{k}^{\dagger }\lambda _{k}),
\end{equation}%
where $\alpha ,\beta ,\gamma $, and $\lambda $ denote four excitation
spectra with the dispersion relations $\omega _{k}^{i}$ ($i$=$\alpha ,\beta
,\gamma ,\lambda $).

\subsection{Spin gap and magnetization}

\begin{figure}[tbp]
\includegraphics[width=1.0\linewidth,clip]{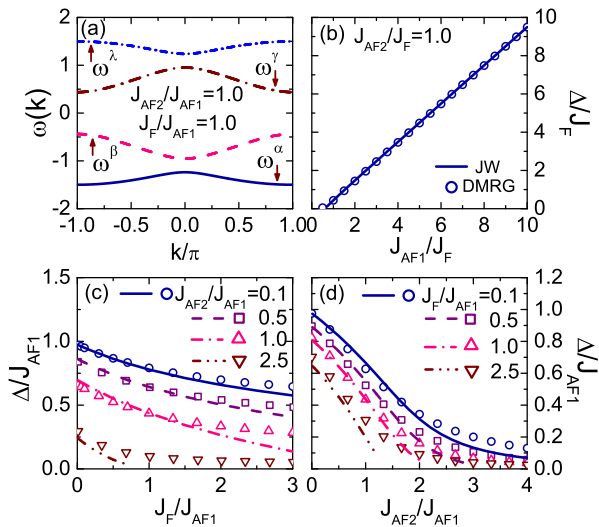}
\caption{(Color online) (a) The elementary excitation spectra of the system.
The gap as a function of (b) $J_{\mathrm{AF_{1}}}$; (c) $J_{\mathrm{F}}$;
and (d) $J_{\mathrm{AF_{2}}}$. The DMRG results are from Ref. 
\onlinecite{GSS}.}
\label{ground}
\end{figure}

In the mean-field calculations of the ground-state properties, the
occupation numbers and covalent bondings are self-consistently calculated by
minimizing the ground state energy with the constraint $\langle S_{\mathrm{%
tot}}^{z}\rangle$=$\langle \sum_{i=1}^{4N}S_{i}^{z}\rangle $=$0$, i.e., the
total number of spinless fermions is $2N$. The self-consistent calculations
give rise to four excitation spectra [Fig. \ref{ground}(a)]. In the absence
of magnetic field, $\omega _{k}^{\alpha }$=$-\omega _{k}^{\lambda }$ and $%
\omega _{k}^{\beta }$=$-\omega _{k}^{\gamma }$. Thus, the ground state is
obtained by filling up the two negative spectra $\alpha $ and $\beta $. The
gap from $S_{\mathrm{tot}}^{z}$=$0$ to $S_{\mathrm{tot}}^{z}$=$\pm 1$,
corresponds to the energy of adding or removing a fermion 
\begin{eqnarray}
\Delta _{HF} &=&\omega _{k=\pi }^{\gamma }=\frac{\sqrt{2}}{4}%
\{A-\{A^{2}-4[J_{\mathrm{AF_{1}}}^{4}(1+2p_{AB})^{2}  \nonumber \\
&\times&(1+2p_{CD}^{2})-2J_{\mathrm{AF_{1}}}^{2}J_{\mathrm{AF_{2}}}J_{%
\mathrm{F}}(1+2p_{AB})  \nonumber \\
&\times &(1+2p_{BC})(1+2p_{CD})(1+2p_{DA})  \nonumber \\
&+&J_{\mathrm{AF_{2}}}^{2}J_{\mathrm{F}}^{2}(1+2p_{BC})^{2}(1+2p_{DA})^{2}]%
\}^{\frac{1}{2}}\}^{\frac{1}{2}},
\end{eqnarray}%
where the parameter $A$ is 
\begin{eqnarray}
A &=&J_{\mathrm{AF_{1}}}^{2}[(1+2p_{AB})^{2}+(1+2p_{CD})^{2}]  \nonumber \\
&+&J_{\mathrm{AF_{2}}}^{2}(1+2p_{BC})^{2}+J_{\mathrm{F}}^{2}(1+2p_{DA})^{2}.
\end{eqnarray}%
For $J_{\mathrm{AF_{1}}}$=$J_{\mathrm{AF_{2}}}$=$J_{\mathrm{F}}$=$1$, the
mean-field result gives the gap $\Delta _{HF}$=$0.4377J_{\mathrm{AF_{1}}}$,
which agrees with the DMRG result $\Delta$=$0.435J_{\mathrm{AF_{1}}}$.\cite%
{GSS}

In Fig. \ref{ground}(b), the gap is plotted as a function of $J_{\mathrm{%
AF_{1}}}$. The HF results agree with the DMRG values when $J_{\mathrm{AF_{1}}%
}/J_{\mathrm{F}}$$>$$0.7$. In Figs. \ref{ground}(c) and (d), $J_{\mathrm{F}}$
and $J_{\mathrm{AF_{2}}}$ dependences of the gap are plotted, respectively.
It can be seen that when $J_{\mathrm{F}}$ and $J_{\mathrm{AF_{2}}}$ are
smaller than $J_{\mathrm{AF_{1}}}$, the HF results are consistent with the
DMRG results, but with increasing $J_{\mathrm{F}}$ or $J_{\mathrm{AF_{2}}}$,
the HF results become worse when $J_{\mathrm{F}}$ or $J_{\mathrm{AF_{2}}}$
is prominent.

In a magnetic field, new quasiparticles can be excited in the $\gamma $ and $%
\lambda $ branches. It is found that the excited fermions can interpret the
magnetization process at zero temperature. When the field is less than the
gap, no fermion is excited and $m$=$0$. When the field closes the gap at $%
h_{c_{1}}$, new fermions are excited in the $\gamma $ branch, and $m$
increases until the $\gamma $ branch is fully filled at $h_{c_{2}}$. The gap
between the spectra $\gamma $ and $\lambda $ is the width of the plateau,
say, $h_{c_{3}}-h_{c_{2}}$. When the field exceeds $h_{c_{3}}$, the fermions
in the spectrum $\lambda $ are excited, and the system is fully
spin-polarized.

\subsection{Zero-field thermodynamics}

At finite temperature, the numbers of the excitations obey the Fermi
distribution function $n_{k}^{i}=1/(e^{\omega _{k}^{i}/T}+1)$ ($i=\alpha
,\beta ,\gamma ,\lambda $), from which the thermodynamics can be obtained by
the self-consistent calculations for the occupation numbers and covalent
bondings.

\begin{figure}[tbp]
\includegraphics[width=1.0\linewidth,clip]{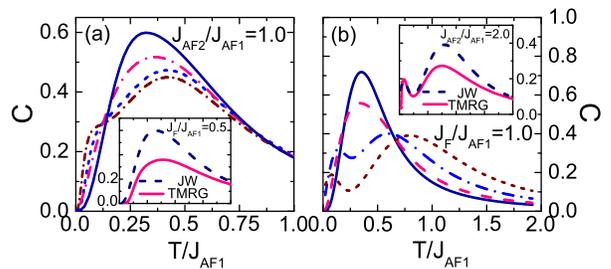}
\caption{(Color online) Temperature dependence of the specific heat for (a) $%
J_{\mathrm{F}}$/$J_{\mathrm{AF_{1}}}$=$0.5,1.5,2.5$ and $3.0$; and (b) $J_{%
\mathrm{AF_{2}}}$/$J_{\mathrm{AF_{1}}}$=$0.5,1.0,1.5$ and $2.0$ from top to
bottom. The comparisons with the TMRG results are shown in the insets.}
\label{thermodynamics}
\end{figure}

In Fig. \ref{thermodynamics}, the specific heat obtained from the HF
calculations are displayed, which qualitatively agree with the TMRG results.
With increasing $J_{\mathrm{F}}$ or $J_{\mathrm{AF_{2}}}$, the gap is
decreased. Therefore, the peaks of the specific heat contributed from $\beta 
$ and $\gamma $ spectra move to lower temperatures with the height
decreased, while those from $\alpha $ and $\lambda $ spectra shift to higher
temperatures with the height declined, both of which compose the observed
shoulder and double peaks in the specific heat. As the spectra $\beta $ and $%
\gamma $ shift slowly with $J_{\mathrm{F}}$ but rapidly with $J_{\mathrm{%
AF_{2}}}$, the specific heat exhibits a shoulder with increasing $J_{\mathrm{%
F}}$ but double peaks with increasing $J_{\mathrm{AF_{2}}}$ at low
temperature. The three peaks in the specific heat that cannot be reproduced
in the mean-field theory may result from the interactions of the
quasiparticles, which might induce excitations between $\alpha $($\lambda $)
and $\beta $($\gamma $). It is expected that the energies of the induced
excitations are enhanced with increasing $J_{\mathrm{F}}$, yielding the
observed shift of the novel peak from the low to high temperatures in Fig. %
\ref{Cv2}(b).

\subsection{Thermodynamics in magnetic fields}

At zero temperature, the spectra $\alpha $ and $\beta $ are fully occupied,
while $\gamma $ and $\lambda $ spectra are empty in the absence of magnetic
field. Thus, the excitations at finite temperature are of the hole ($\alpha $
and $\beta $) and fermion ($\gamma $ and $\lambda $) types, respectively. In
a magnetic field, the energies of the spectra decrease, yielding the
excitations with minus energies to become hole excitations at finite
temperature. In this subsection, some typical thermodynamic behaviors could
be explained by the combination of the hole and fermion excitations.

\begin{figure}[tbp]
\includegraphics[width=1.0\linewidth,clip]{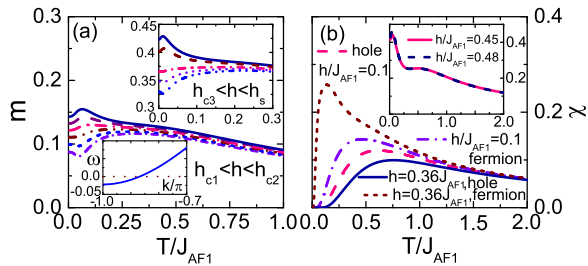}
\caption{(Color online) Mean-field results in different magnetic fields for
(a) magnetization $m(T)$ at low temperatures in the gapless regime; (b)
susceptibility contributed from the fermion and hole excitations. The lower
inset of (a) depicts the dispersion in the vicinity of the lower critical
field.}
\label{thermodynamics2}
\end{figure}

When $h_{c_{1}}$$<$$h$$<$$h_{c_{2}}$ and $h_{c_{3}}$$<$$h$$<$$h_{s}$, the
minimum and maximum of $m(T)$ at low temperature are observed in the
mean-field results, and the crossover temperatures in the gapless regime are
lower than those in the LL, as shown in Fig. \ref{thermodynamics2}(a). As
the magnetization is proportional to the total number of the fermions, the
minimum and maximum indicate that the hole and fermion excitations dominate
at low temperature, respectively. In the vicinity of the lower critical
field $h_{c_{1}}$ ($h_{c_{3}}$), the fermion spectrum $\gamma $ ($\lambda $)
crosses the Fermi level slightly, as shown in the lower inset of Fig. \ref%
{thermodynamics2}(a), making the excitations with minus energies become hole
type. Thus, the low-temperature behavior of $m(T)$ is determined by the
competitions of the fermion and hole excitations near the Fermi level. Due
to the dispersion relation, the density of states $D(\omega )$ of the holes
is larger than that of the fermions near the Fermi level, yielding more
holes excited and thus the decrease of $m$ at low temperature. After the few
holes are occupied, the fermion excitations dominate and, thus, a minimum of 
$m$ emerges. The analogous arguments for the maximum near the upper critical
field $h_{c_{2}}$ ($h_{s}$) also apply. The lower crossover temperature in
the gapless regime is attributed to the larger values of the gradient of the
Fermi velocity with respect to the field $\partial v_{F}$/$\partial h$ of
the spectrum $\lambda $ near the Fermi level, which diminishes the
difference of the density of states in the vicinity of the Fermi level.

Different from the magnetization, the fermion and hole excitations have
equivalent contributions to the susceptibility. However, with increasing the
field, the fermion excitations move close to the Fermi surface, while the
holes move away. Thus, the peak of the susceptibility from the excited
fermions shifts to lower temperatures with the height enhanced, while that
from the excited holes exhibits opposite behaviors, as shown in Fig. \ref%
{thermodynamics2}(b). All the observed behaviors of the susceptibility in
the gapped spin liquid and plateau phases can be understood on the basis of
the above analyses, as mentioned in Sec. \uppercase\expandafter{%
\romannumeral3}.

Now let us discuss breifly the specific heat. The specific heat can be
divided into those from the fermion and hole excitations, which are denoted
as $C_{f}$ and $C_{h}$, respectively. It is found that $C_{f}$ and $C_{h}$
behave rather differently with the field, and $C_{h}$ contributes
essentially to the observed characteristic features, which are not shown
here for brevity. For example, when $h$$<$$h_{c_{1}}$, the gaps of the hole
excitations $\alpha $ and $\beta $ are enhanced with increasing the field,
making the peaks of $C_{h}$ move to higher temperatures with the amplitude
decreased. As $h$ approaches $h_{c_{1}}$, a low temperature peak in $C_{h}$
emerges, yielding the shoulder in the specific heat. In other phases,
similar analyses are also applicable.

As discussed above, the spinless fermion mapping can be used to explain both
the ground-state and thermodynamic properties in a wide range of the
parameters. The fermion and hole excitations provide a simple way to
understand the complex thermodynamic behaviors of the system.

\section{Summary and Discussion}

The thermodynamic properties of the spin $S$=$1/2$ tetrameric HAFC with
alternating couplings AF$_{1}$-AF$_{2}$-AF$_{1}$-F have been studied by
means of the TMRG method and JW transformation. In the absence of magnetic
field, the thermodynamic behaviors are determined by the gapped low-lying
excitations. The specific heat can have single peak, double peaks, and three
peaks for different couplings. The shoulder and double peaks in the specific
heat are attributed to the decreased energies of the gapped excitations. The novel
intermediate peak in the specific heat results from the increase of $J_{%
\mathrm{F}}$ after the double peaks have been induced by large $J_{\mathrm{%
AF_{2}}}$. With further increase of $J_{\mathrm{F}}$, the novel peak shifts
to higher temperatures and finally merges into the high-temperature peak.
The susceptibility is found to have a peak that shifts to lower temperatures
with the amplitude enhanced with increasing $J_{\mathrm{F}}$ or $J_{\mathrm{%
AF_{2}}}$.

A phase diagram in the temperature-field plane for $J_{\mathrm{AF_{1}}}$=$J_{%
\mathrm{AF_{2}}}$=$J_{\mathrm{F}}$ is obtained by analyzing the crossover
behaviors of the various phases at finite temperature. The system is
unveiled to contain the gapped spin liquid, LL, magnetization plateau,
gapless, spin polarized, and classical phases.

In the LL and gapless regimes, the magnetization curve exhibits a minimum or
maximum at low temperature, representing a nonsingular crossover. The linear
field dependence of the crossover temperature in the LL with the same ratio
as that in the $S$=$1$ Haldane chain indicates that the crossovers in the
two systems belong to the same class, which implies that the Haldane-like
phase of this tetrameric HAFC has an analogous low-lying dispersion relation
to that in the $S$=$1$ Haldane chain. In the plateau state, as a crossing of
a fermion and a hole spectra at $h_{m}$, the temperature dependence of the
magnetization behaves differently below and above the field.

When $h$$<$$h_{c_{1}}$, the susceptibility has double peaks with increasing
the field. In the plateau state, the susceptibility has a single peak, and
owing to the crossing of the spectra, the peak moves to higher temperatures
when $h$$<$$h_{m}$, and to lower temperatures when $h$$>$$h_{m}$ with
increasing the field. In the LL and gapless regimes, $\chi $ is finite as $T$%
$\rightarrow $$0$. The susceptibility in any fields coincides when $T$$>$$%
\Delta$ due to the thermal fluctuations.

In the gapped spin liquid, the specific heat has a shoulder as the field
approaches the critical field $h_{c_{1}}$. In the LL, the specific heat as a
function of temperature behaves linearly at low temperature, and the
shoulder is smoother down gradually by increasing the field. In the plateau,
owing to the crossing of spectra, the peak of the specific heat moves to
lower temperatures when $h$$<$$h_{m}$, and to higher temperatures when $h$$>$%
$h_{m}$ with increasing the field. As the field approaches $h_{c_{3}}$, a
small peak emerges at low temperatures. In the gapless phase, with increased field
$h-h_{c_{3}}$, the low-temperature peak moves to lower temperature with the amplitude decreased.

By means of the JW transformation and mean-field approximation, the
low-lying excitations and thermodynamic properties of the system are studied
to understand the TMRG results. It is unveiled that the system has four
excitation spectra with a gap, which may account for the magnetization
process at zero temperature. At finite temperature, it is found that the
thermodynamic behaviors are determined by the combination of the fermion and
hole excitations. The complex thermodynamic behaviors in the TMRG results
can be understood well within the free fermion mapping.

Finally, we would like to add that the observations presented in this paper
for this spin-1/2 tetrameric HAFC would be expected to test experimentally
in future.

\acknowledgments

We are grateful to Bo Gu, Wei Li, Yang Zhao, and Guang-Qiang Zhong for
useful discussions. This work is supported in part by the National Science
Fund for Distinguished Young Scholars of China (Grant No. 10625419), the
MOST of China (Grant No. 2006CB601102), and the Chinese Academy of Sciences.

\end{document}